\RequirePackage{fix-cm}
\documentclass[smallextended]{svjour3}       
\smartqed  
\usepackage{graphicx}
\usepackage{subfigure}
\begin{document}

\title{Smooth positon solutions of the focusing modified Korteweg-de Vries equation  \thanks{Corresponding author: hejingsong@nbu.edu.cn; jshe@ustc.edu.cn}
}
\subtitle{}


\author{Qiuxia Xing       \and
        Zhiwei Wu \and Dumitru Mihalache \and  Jingsong He
}


\institute{Qiuxia Xing, Zhiwei Wu, Jingsong He \at
              Department of Mathematics,  Ningbo University, Zhejiang 315211, P. R. China \\
              Tel.: +86 574-87600739\\
              Fax: +86 574-87600744\\
              \email{hejingsong@nbu.edu.cn;jshe@ustc.edu.cn}           
           \and
           Dumitru Mihalache \at
              Horia Hulubei National Institute for Physics and Nuclear Engineering, P.O.Box MG--6, Magurele, 077125, Romania
}

\date{Received: date / Accepted: date}

\maketitle

\begin{abstract}
The
$n$-fold Darboux transformation $T_{n}$ of the focusing  real mo\-di\-fied Kor\-te\-weg-de Vries (mKdV) equation is expressed in terms of the determinant representation. Using this representation, the $n$-soliton solutions of the mKdV equation are also  expressed by determinants whose elements consist of  the eigenvalues $\lambda_{j}$ and the corresponding eigenfunctions of the associated Lax equation.  The nonsingular $n$-positon solutions of the focusing  mKdV equation are obtained in the special limit $\lambda_{j}\rightarrow\lambda_{1}$, from the corresponding $n$-soliton solutions and by using the associated higher-order Taylor expansion. Furthermore, the decomposition method of the $n$-positon solution into $n$
single-soliton solutions, the trajectories, and the corresponding ``phase shifts" of the multi-positons are also investigated.
\end{abstract}

\section{Introduction}
\label{intro}
It is a well-known fact that nonlinear partial differential equations play a fundamental role both in the understanding of many natural phenomena and in the development of advanced technologies and engineering designs. A plethora of such nonlinear evolutions equations have been investigated during the many years such as the celebrated Korteweg-de Vries (KdV) equation, the modified Korteweg-de Vries (mKdV) equation, the sine-Gordon (sG) equation, the nonlinear Schr\"{o}dinger (NLS) equation, the Manakov system, Kadomtsev-Petviashvili equation, Davey-Stewartson equation, Maccari system etc. Especially, the origin of the KdV equation and its birth has been a long process and spanned a period of about sixty years from the experiments of Scott-Russell in 1834 \cite{su-1} to the publication in 1895 of a seminal article by Korteweg and de Vries \cite{NK} who developed a mathematical model for the shallow water problem and demonstrated the possibility of solitary wave generation. The KdV equation has been derived in a series of physical settings, e.g. in plasma physics \cite{E-2,E-1}, and in studies of anharmonic (nonlinear) lattices \cite{o-3,o-4}. Here we recall that the existence and uniqueness of solutions of the KdV equation for appropriate initial and boundary conditions have been proved by Sj\"{o}berg \cite{o-5}. It is well known that if $u$ is a solution of the KdV equation $u_{t}+u_{xxx}-6uu_{x}=0$ and $v$ is a solution of the defocusing mKdV equation $v_{t}+v_{xxx}-6v^{2}v_{x}=0$, the two solutions are connected by the Miura transformation \cite{AK1}, namely, $u=v_{x}+v^{2}$. Both KdV and mKdV equations are completely integrable and have infinitely many conserved quantities \cite{AK2}.

The KdV and mKdV equations and its many generalizations were used to describe a series of physical phenomena. For example, the system of coupled KdV equations is a generic model of resonantly coupled internal waves in stratified fluids and can also describe the formation of gap solitons and of parametric envelope solitons, see Ref. \cite{B1}-\cite{B5}.
In optical settings the mKdV equation and its further generalizations were found to adequately describe the propagation in nonlinear optical media of ultrasort pulses consisting of only a few optical cycles, beyond the so-called slowly varying envelope aproximation \cite{LM2009}-\cite{M2015}. Recently, a model based on two coupled mKdV equations was used to describe the soliton propagation in two parallel optical waveguides, in the presence of linear nondispersing coupling and in the few-cycle regime \cite{Terniche2016}. The mKdV, sG, and mKdV-sG equations were used in modelling the process of generation of supercontinuum light in optical fibers \cite{SCG2014,SCG2016}. The mKdV equation also appears in other fields such as ion acoustic soliton experiments in plasmas \cite{YOC1,ion1}, fluid mechanics \cite{RW1}, soliton propagation in lattices and acoustic waves in certain anharmonic lattices \cite{H1}, nonlinear van Alfv\'{e}n waves propagating in plasmas \cite{H2,H22}, meandering ocean currents \cite{H3},
the dynamics of traffic flow \cite{on,Y.11}, and the study of Schottky barrier transmission lines \cite{Y.12}. For a series of recent studies of the KdV and mKdV equations, and other relevant nonlinear partial differential equations that describe the dynamics of diverse physical phenomena, see Refs. \cite{Ra}-\cite{Rk}.  It is worth mentioning that being a completely integrable nonlinear dynamical system, the generic mKdV partial differential equation possesses unique features such as the Painlev\'{e} property \cite{property1,property11}, Miura transformation \cite{property2}, the inverse scattering transformation \cite{property3}, the Darboux transformation (DT)\cite{YO-6} and so on.

In a pioneering work \cite{kdv1992}, Matveev introduced the concept of a {\it positon solution} as a new type of solution of the KdV equation. Both positon and soliton-positon solutions of the KdV equation were first constructed and analyzed \cite{kdv1992}. The positon solutions have many interesting properties that differ from those of soliton solutions. The positons can be characterized as slowly decaying oscillating solutions of the nonlinear completely integrable equations having the special property of being superreflectionless \cite{HE2}. The positons are weakly localized, in contrast to exponentially decaying soliton solutions. For a positon solution the corresponding eigenvalue of the spectral problem is positive and is embedded in the continuous spectrum. The positons are completely transparent to other interacting objects. In particular, two positons remain unchanged after mutual collision. However, during the soliton-positon collision the soliton remains unchanged, while both the carrier-wave of the positon and its envelope experience finite phase-shifts \cite{HE3,onorato}. The positon solutions were then constructed for many other models, such as the defocusing mKdV equation \cite{positon1992,positon1995}, the sG  equation \cite{Y.Ohta12} and the Toda-lattice \cite{YOC}.  It is well known that that the above mentioned positon solutions are singular ones.   So it is important from the point of view of possible applications of positon solutions to model diverse  physical phenomena to seek smooth, nonsingular positon solutions for nonlinear evolution equations.
It is worth noting that it is not possible to get smooth, nonsingular positon solutions of the defocusing mKdV because of the singularity that is inherent in its Darboux transformation. Instead, it is natural to seek
nonsingular positon solutions of the following focusing-type mKdV equation
\begin{equation}\label{MKDV}
q_t=\frac{q_{xxx}}{4}+\frac{3q^2q_x}{2}
\end{equation}
by using the DT method.  Here $q=q(x,t)$ is a real function of  variables $x$ and $t$.  Furthermore, we  know that the multi-soliton solution for the mKdV equation has been given, e.g.,  in Refs. \cite{dubardthesis2010,matveev2011nlsKPI,Gaillard-2011-435204}. We should note that the $n$-soliton solution of the mKdV equation can be decomposed into a summation of $n$ single solitons with a constant phase shift at $|t|\longrightarrow\infty$, see, for example Ref. \cite{matveev2011nlsKPI}. It is the main aim of this paper to decompose the multi-positon solution of the focusing real mKdV equation, and to further study its key properties of the nonsigular positons including their trajectories and the associated ``phase shift".

The organization of this paper is as follows. In Section 2, the $n$-fold Darboux transformation is given. In Section 3, the positon solution of the focusing real mKdV equation is calculated by using the degenerate Darboux transformation method and a  special Taylor expansion. The multi-positon solution and its decomposition into several single-positon solutions, the associated positon trajectories and phase shifts are also discussed. Our conclusions are provided in the last section.

\section{The Darboux transformation of the focusing  mKdV equation}
\label{sec:1}
The Lax pair of the mKdV equation (\ref{MKDV}) can be derived by the Lie algebra splitting \cite{AKNS,DS84,TUa,TUb}:
\begin{equation}\label{5}
\phi_x{(\lambda)}=M\phi{(\lambda)},
\end{equation}
\begin{equation}\label{6}
\phi_t{(\lambda)}=(N_3{\lambda}^3+N_2{\lambda}^2+N_1{\lambda}+N_0)\phi{(\lambda)}=N\phi{(\lambda)},
\end{equation}
with
$$
{\phi(\lambda)}=\left[\begin{array}{ccc}
\varphi_{1}(\lambda)\\
\varphi_{2}(\lambda)
\end{array}\right], \qquad   \qquad
{M}=\left[\begin{array}{ccc}
\lambda &  q \\
-q & -\lambda
\end{array}\right],  \qquad   \qquad
{N_3}=\left[\begin{array}{ccc}
1 &  0 \\
0 &  -1
\end{array}\right],$$
$$
{N_2}=\left[\begin{array}{ccc}
0 &  q \\
-q &  0
\end{array}\right],\qquad
{N_1}=\left[\begin{array}{ccc}
\frac{q^2}{2} &  \frac{q_x}{2}\\
\frac{q_x}{2} & -\frac{q^2}{2}
\end{array}\right],\qquad
{N_0}=\left[\begin{array}{ccc}
0 &  \frac{q_{xx}+2q^3}{4}\\
\frac{-q_{xx}-2q^3}{4} & 0
\end{array}\right].
$$

Here $\phi(\lambda)$ is an eigenfunction of the Lax equation associated with eigenvalue $\lambda$.
In order to later construct the kernel of  $n$-fold DT, it is necessary to introduce 2n eigenfunctions:
$$\phi_{2j-1}=\phi{(\lambda)}|_{\lambda=\lambda_{2j-1}}=\left[\begin{array}{c}
\phi_{2j-1,1}\\
\phi_{2j-1,2}\\
\end{array}\right]  \quad
\phi_{2j}=\phi{(\lambda)}|_{\lambda=\lambda_{2j}}=\left[\begin{array}{c}
\phi_{2j,1}\\
\phi_{2j,2}\\
\end{array}\right].
$$
The even-numbered functions are given through the reduction condition:\\
\begin{equation}\label{reduction}
\lambda_{2j}=-\lambda_{2j-1}, \quad
\phi_{2j}=\left[\begin{array}{cc}
\phi_{2j,1}\\
\phi_{2j,2}
\end{array}\right]
=
\left[\begin{array}{c}
-\phi_{2j-1,2}\\
\phi_{2j-1,1}
\end{array}\right].
\end{equation}

Here  $\varphi_{1}(x,t,\lambda)$ and $\varphi_{2}(x,t,\lambda)$ are the two components of eigenfunction $\phi$ associated with $\lambda$ in Eqs. (\ref{5}) and (\ref{6}). Furthermore, we will set $T$ to be a gauge transformation as follows
\begin{equation}\label{7}
\phi^{[1]}=T\phi,~~q\rightarrow q^{[1]},
\end{equation}
\begin{equation}\label{8}
\phi_x^{[1]}=M^{[1]}\phi^{[1]},
\end{equation}
\begin{equation}\label{9}
\phi_t^{[1]}=N^{[1]}\phi^{[1]}.
\end{equation}

According to Eqs. (\ref{5}), (\ref{6}), (\ref{7}), (\ref{8}), and (\ref{9}), it is easy to get
\begin{equation}\label{xpartDT}
M^{[1]}T=T_x+TM.
\end{equation}
Similarly, the following equation can be also obtained
\begin{equation}\label{11}
N^{[1]}T=T_t+TN.
\end{equation}
Hence, we get
\begin{equation}\label{12}
M_t^{[1]}-N_x^{[1]}+[M^{[1]},N^{[1]}]=0.
\end{equation}

In general, in order to seek for Darboux transformation, we assume
\begin{equation}
{T}=\left[\begin{array}{ccc}
a_1 &  b_1 \\
c_1 &  d_1
\end{array}\right]\lambda+\left[\begin{array}{ccc}
a_0 &  b_0 \\
c_0 &  d_0
\end{array}\right].
\end{equation}
Here, $a_0, b_0, c_0, d_0, a_1, b_1, c_1$, and $d_1$ are functions of $x$ and $t$, which can be easily obtained according to
the Eqs. (\ref{11}) and (\ref{12}), by comparing the coefficients of $\lambda^i \, (i=4, 3, 2, 1, 0)$.

For example,  it is easily  to get $a_{1t}=0,\quad b_{1t}=0$,
by comparing the coefficients of $\lambda$. Furthermore,
$a_{1x}=0,\quad b_{1x}=0,$
so $a_{1}$ and $d_{1}$ are arbitrary constants. Then we  set $a_1=1$,~~$b_1=1$ in order to simplify the later calculations without loss of generality, i.e.,
\begin{equation}
{T_{1}}=\left[\begin{array}{ccc}
1 &  0 \\
0 &  1
\end{array}\right]\lambda+\left[\begin{array}{ccc}
a_0 &  b_0 \\
c_0 &  d_0
\end{array}\right].
\end{equation}

Then, $a_0$, ~$b_0$,~$c_0$, and ~$d_0$ can be obtained by using the kernel of $T_1$:
$$T_{1}(\lambda; \lambda_{1}, \lambda_{2})\phi{(\lambda)}|_{\lambda=\lambda_{1}}=0, \quad
T_{1}(\lambda; \lambda_{1}, \lambda_{2})\phi{(\lambda)}|_{\lambda=\lambda_{2}}=0.$$
Here
the two eigenfunctions $\phi{(\lambda_{1})}$ and $\phi{(\lambda_{2})}$ are in the form of
$$\phi(\lambda_{1})=\left[\begin{array}{ccc}
\phi_{11}\\
\phi_{12}
\end{array}\right], \qquad
\phi(\lambda_{2})=\left[\begin{array}{ccc}
\phi_{21}\\
\phi_{22}
\end{array}\right].$$

Solving the above algebraic equations, then
$$
{a_{0}}=-\frac{\left|\begin{array}{ccc}
        \lambda_1\phi_{11} &  \phi_{12}\\
        \lambda_2\phi_{21} &  \phi_{22}
      \end{array}\right|}{|W_{2}|}, \qquad
{b_{0}}=\frac{\left|\begin{array}{ccc}
        \lambda_1\phi_{11} &  \phi_{11}\\
        \lambda_2\phi_{21} &  \phi_{21}
      \end{array}\right|}{|W_{2}|},
$$
$$
{c_{0}}=\frac{\left|\begin{array}{ccc}
        \phi_{12} & \lambda_1\phi_{12} \\
         \phi_{22} & \lambda_2\phi_{21}
\end{array}\right|}{|W_{2}|},   \qquad
{d_{0}}=-\frac{\left|\begin{array}{ccc}
        \phi_{11} & \lambda_1\phi_{12} \\
          \phi_{21} & \lambda_2\phi_{22}
      \end{array}\right|}{|W_{2}|}.
$$

Taking these elements back into $T_1$, then the determinant representation of $T_1$ is given by
\begin{equation}
{T_{1}}=\frac{1}{|W_{2}|}{\left|\begin{array}{ccccc}
(\widetilde{T_{1}})_{11}& (\widetilde{T_{1}})_{12}\\
(\widetilde{T_{1}})_{21}&  (\widetilde{T_{1}})_{22}
\end{array}\right|}, \qquad   \qquad
{W_{2}}=\left[\begin{array}{ccc}
\phi_{11} &  \phi_{12} \\
\phi_{21} &  \phi_{22}
\end{array}\right],
\end{equation}
$$
{(\widetilde{T_{1}})_{11}}=\frac{\left|\begin{array}{ccccc}
1         &  0         &\lambda         \\
\phi_{11} &  \phi_{12} &  \lambda_{1}\phi_{11} \\
\phi_{21} &  \phi_{22} &  \lambda_{2}\phi_{21}
\end{array}\right|}{|W_{2}|},    \qquad
{(\widetilde{T_{1}})_{12}}=\frac{\left|\begin{array}{ccccc}
0         &  1        &            0 \\
\phi_{11} &  \phi_{12} &  \lambda_{1}\phi_{11} \\
\phi_{21} &  \phi_{22} &  \lambda_{2}\phi_{21}
\end{array}\right|}{|W_{2}|},
$$
$$
{(\widetilde{T_{1}})_{21}}=\frac{\left|\begin{array}{ccccc}
1         &         0 &            \lambda \\
\phi_{11} &  \phi_{12} &  \lambda_{1}\phi_{11} \\
\phi_{21} &  \phi_{22} &  \lambda_{2}\phi_{21}
\end{array}\right|}{|W_{2}|},   \qquad
{(\widetilde{T_{1}})_{22}}=\frac{\left|\begin{array}{ccccc}
0         &  1        &            0 \\
\phi_{11} &  \phi_{12} &  \lambda_{1}\phi_{11} \\
\phi_{21} &  \phi_{22} &  \lambda_{2}\phi_{21}
\end{array}\right|}{|W_{2}|}.
$$

Further, the
one-fold Darboux transformation $T_1$ generates
a new solution of the focusing mKdV equation
\begin{equation}
{q^{[1]}}=q+\frac{2}{|W_{2}|}{\left|\begin{array}{ccc}
\phi_{11}&  \lambda_1\phi_{11}\\
\phi_{21} & \lambda_2\phi_{21}
\end{array}\right|},
\end{equation}
according to Eq. (\ref{xpartDT}). Note that $\phi_2$ and $\lambda_2$ are given by the reduction
condition in Eq. (\ref{reduction})

  By iterating $n$ times $T_1$ and using the representation of $T_1$, it is easy to get the
$n$-fold Darboux transformation, see the following theorem. \\
\textbf{Theorem} The $n$-fold Darboux transformation of the mKdV equation (\ref{MKDV}) is as follows
\begin{equation}
{T_{n}}=T_{n}(\lambda)=\frac{\left|\begin{array}{ccccc}
(\widetilde{T_{n}})_{11}& (\widetilde{T_{n}})_{12}\\
(\widetilde{T_{n}})_{21}&  (\widetilde{T_{n}})_{22}
\end{array}\right|}{|W_{2n}|},
\end{equation}
where
{\footnotesize
$${W_{2n}}=\left[\begin{array}{ccccccc}
\phi_{11} &  \phi_{12} &\lambda_{1}\phi_{11}&  \lambda_{1}\phi_{12}&   \cdots\lambda_{1}^{n-1}\phi_{11}&  \lambda_{1}^{n-1}\phi_{12}\\
\phi_{21} &  \phi_{22} &\lambda_{2}\phi_{21}&  \lambda_{2}\phi_{22}&   \cdots\lambda_{2}^{n-1}\phi_{11}&  \lambda_{2}^{n-1}\phi_{12}\\
\phi_{31} &  \phi_{32} &\lambda_{3}\phi_{31}&  \lambda_{3}\phi_{32}&   \cdots\lambda_{3}^{n-1}\phi_{11}&  \lambda_{3}^{n-1}\phi_{12}\\
\vdots    &  \vdots    &\vdots              &  \vdots             &               \vdots              &  \vdots\\
\phi_{2n,1} &  \phi_{2n,2} &\lambda_{2n}\phi_{2n,1}&  \lambda_{2n}\phi_{2n,2}&   \lambda_{2n}^{n-1}\phi_{2n,1}&  \lambda_{2n}^{n-1}\phi_{2n,2}
\end{array}\right],$$
}
{\footnotesize
$$\mbox{\hspace{-0.2cm}}\mathbf{(\widetilde{T_n})_{11}}=\frac{\left|\begin{array}{cccccccccc}
1         &  0         &\lambda             &  0                   &    \lambda^2         & 0                     &  \cdots     \lambda^{n-1}         &  0                                &\lambda^n\\
\phi_{11} &  \phi_{12} &\lambda_{1}\phi_{11}&  \lambda_{1}\phi_{12}& \lambda_{1}^2\phi_{11}&\lambda_{1}^2\phi_{12}&  \cdots\lambda_{1}^{n-1}\phi_{11}& \lambda_{1}^{n-1}\phi_{12}& \lambda_{1}^{n}\phi_{11}\\
\phi_{21} &  \phi_{22} &\lambda_{2}\phi_{21}&  \lambda_{2}\phi_{22}& \lambda_{2}^2\phi_{21}&\lambda_{2}^2\phi_{22}&  \cdots\lambda_{2}^{n-1}\phi_{21}& \lambda_{2}^{n-1}\phi_{22}& \lambda_{2}^{n}\phi_{21}\\
\phi_{31} &  \phi_{32} &\lambda_{3}\phi_{31}&  \lambda_{3}\phi_{32}& \lambda_{3}^2\phi_{31}&\lambda_{3}^2\phi_{32}&  \cdots\lambda_{3}^{n-1}\phi_{31}& \lambda_{3}^{n-1}\phi_{32}& \lambda_{3}^{n}\phi_{31}\\
\vdots    &  \vdots    &\vdots              &  \vdots             &               \vdots              &  \vdots  &  \vdots             &               \vdots              &  \vdots\\
\phi_{2n,1} &  \phi_{2n,2} &\lambda_{n}\phi_{2n,1}&  \lambda_{n}\phi_{2n,2}& \lambda_{n}^2\phi_{2n,1}&\lambda_{n}^2\phi_{2n,2}&  \cdots\lambda_{n}^{n-1}\phi_{2n,1}& \lambda_{n}^{n-1}\phi_{2n,2}& \lambda_{n}^{n}\phi_{2n,1}
\end{array}\right|}{|W_{2n}|},$$
}
{\footnotesize
$$\mbox{\hspace{-0.2cm}}\mathbf{(\widetilde{T_n})_{12}}=\frac{\left|\begin{array}{cccccccccc}
0         &  1         &0                   &  \lambda             &    0               & \lambda^2               & \cdots     0                     &  \lambda^{n-1}                                 &0\\
\phi_{11} &  \phi_{12} &\lambda_{1}\phi_{11}&  \lambda_{1}\phi_{12}& \lambda_{1}^2\phi_{11}&\lambda_{1}^2\phi_{12}&  \cdots\lambda_{1}^{n-1}\phi_{11}& \lambda_{1}^{n-1}\phi_{12}& \lambda_{1}^{n}\phi_{11}\\
\phi_{21} &  \phi_{22} &\lambda_{2}\phi_{21}&  \lambda_{2}\phi_{22}& \lambda_{2}^2\phi_{21}&\lambda_{2}^2\phi_{22}&  \cdots\lambda_{2}^{n-1}\phi_{21}& \lambda_{2}^{n-1}\phi_{22}& \lambda_{2}^{n}\phi_{21}\\
\phi_{31} &  \phi_{32} &\lambda_{3}\phi_{31}&  \lambda_{3}\phi_{32}& \lambda_{3}^2\phi_{31}&\lambda_{3}^2\phi_{32}&  \cdots\lambda_{3}^{n-1}\phi_{31}& \lambda_{3}^{n-1}\phi_{32}& \lambda_{3}^{n}\phi_{31}\\
\vdots    &  \vdots    &\vdots              &  \vdots             &               \vdots              &  \vdots  &  \vdots             &               \vdots              &  \vdots\\
\phi_{2n,1} &  \phi_{2n,2} &\lambda_{n}\phi_{2n,1}&  \lambda_{n}\phi_{2n,2}& \lambda_{n}^2\phi_{2n,1}&\lambda_{n}^2\phi_{2n,2}&  \cdots\lambda_{n}^{n-1}\phi_{2n,1}& \lambda_{n}^{n-1}\phi_{2n,2}& \lambda_{n}^{n}\phi_{2n,1}
\end{array}\right|}{|W_{2n}|},$$
}
{\footnotesize
$$\mbox{\hspace{-0.2cm}}\mathbf{(\widetilde{T_n})_{21}}=\frac{\left|\begin{array}{cccccccccc}
1         &  0         &\lambda             &  0                   &    \lambda^2         & 0                     &  \cdots\lambda^{n-1}        &  0                                &0\\
\phi_{11} &  \phi_{12} &\lambda_{1}\phi_{11}&  \lambda_{1}\phi_{12}& \lambda_{1}^2\phi_{11}&\lambda_{1}^2\phi_{12}&  \cdots\lambda_{1}^{n-1}\phi_{11}& \lambda_{1}^{n-1}\phi_{12}& \lambda_{1}^{n}\phi_{11}\\
\phi_{21} &  \phi_{22} &\lambda_{2}\phi_{21}&  \lambda_{2}\phi_{22}& \lambda_{2}^2\phi_{21}&\lambda_{2}^2\phi_{22}&  \cdots\lambda_{2}^{n-1}\phi_{21}& \lambda_{2}^{n-1}\phi_{22}& \lambda_{2}^{n}\phi_{21}\\
\phi_{31} &  \phi_{32} &\lambda_{3}\phi_{31}&  \lambda_{3}\phi_{32}& \lambda_{3}^2\phi_{31}&\lambda_{3}^2\phi_{32}&  \cdots\lambda_{3}^{n-1}\phi_{31}& \lambda_{3}^{n-1}\phi_{32}& \lambda_{3}^{n}\phi_{31}\\
\vdots    &  \vdots    &\vdots              &  \vdots             &               \vdots              &  \vdots  &  \vdots             &               \vdots              &  \vdots\\
\phi_{2n,1} &  \phi_{2n,2} &\lambda_{n}\phi_{2n,1}&  \lambda_{n}\phi_{2n,2}& \lambda_{n}^2\phi_{2n,1}&\lambda_{n}^2\phi_{2n,2}&  \cdots\lambda_{n}^{n-1}\phi_{2n,1}& \lambda_{n}^{n-1}\phi_{2n,2}& \lambda_{n}^{n}\phi_{2n,1}
\end{array}\right|}{|W_{2n}|},$$
}
{\footnotesize
$$\mbox{\hspace{-0.2cm}}\mathbf{(\widetilde{T_n})_{22}}=\frac{\left|\begin{array}{cccccccccc}
0         &  1         &0                   &  \lambda             &    0               & \lambda^2               &  \cdots     0        &  \lambda^{n-1}                                 &\lambda^n\\
\phi_{11} &  \phi_{12} &\lambda_{1}\phi_{11}&  \lambda_{1}\phi_{12}& \lambda_{1}^2\phi_{11}&\lambda_{1}^2\phi_{12}&  \cdots\lambda_{1}^{n-1}\phi_{11}& \lambda_{1}^{n-1}\phi_{12}& \lambda_{1}^{n}\phi_{11}\\
\phi_{21} &  \phi_{22} &\lambda_{2}\phi_{21}&  \lambda_{2}\phi_{22}& \lambda_{2}^2\phi_{21}&\lambda_{2}^2\phi_{22}&  \cdots\lambda_{2}^{n-1}\phi_{21}& \lambda_{2}^{n-1}\phi_{22}& \lambda_{2}^{n}\phi_{21}\\
\phi_{31} &  \phi_{32} &\lambda_{3}\phi_{31}&  \lambda_{3}\phi_{32}& \lambda_{3}^2\phi_{31}&\lambda_{3}^2\phi_{32}&  \cdots\lambda_{3}^{n-1}\phi_{31}& \lambda_{3}^{n-1}\phi_{32}& \lambda_{3}^{n}\phi_{31}\\
\vdots    &  \vdots    &\vdots              &  \vdots             &               \vdots              &  \vdots  &  \vdots             &               \vdots              &  \vdots\\
\phi_{2n,1} &  \phi_{2n,2} &\lambda_{n}\phi_{2n,1}&  \lambda_{n}\phi_{2n,2}& \lambda_{n}^2\phi_{2n,1}&\lambda_{n}^2\phi_{2n,2}&  \cdots\lambda_{n}^{n-1}\phi_{2n,1}& \lambda_{n}^{n-1}\phi_{2n,2}& \lambda_{n}^{n}\phi_{2n,1}
\end{array}\right|}{|W_{2n}|}.$$
}

Therefore,  the $n$th-order solution generated by the transformation $T_n$ from a ``seed" solution $q$ is given by
\begin{equation}\label{13}
q^{[n]}=q+2\frac{N_{2n}}{W_{2n}},
\end{equation}
where
$$
{N_{2n}}=\left[\begin{array}{ccccccc}
\phi_{11} &  \phi_{12} &\lambda_{1}\phi_{11}&  \lambda_{1}\phi_{12}&   \cdots\lambda_{1}^{n-1}\phi_{11}&  \lambda_{1}^{n}\phi_{11}\\
\phi_{21} &  \phi_{22} &\lambda_{2}\phi_{21}&  \lambda_{2}\phi_{22}&   \cdots\lambda_{2}^{n-1}\phi_{21}&  \lambda_{2}^{n}\phi_{21}\\
\phi_{31} &  \phi_{32} &\lambda_{3}\phi_{31}&  \lambda_{3}\phi_{32}&   \cdots\lambda_{3}^{n-1}\phi_{31}&  \lambda_{3}^{n}\phi_{31}\\
\vdots    &  \vdots    &\vdots              &  \vdots             &               \vdots              &  \vdots\\
\phi_{2n,1} &  \phi_{2n,2} &\lambda_{2n}\phi_{2n,1}&  \lambda_{2n}\phi_{2n,2}&   \lambda_{2n}^{n-1}\phi_{2n,1}&  \lambda_{2n}^{n}\phi_{2n,1}
\end{array}\right].
$$

\section{The $n$-positon solutions of the focusing mKdV equation}
\label{sec:2}
In the previous Section, we have derived the expression of the $n$-fold DT and in what follows we will give the soliton  solution of the mKdV equation in order to study the associated positon solution.  To this end, we set the seed solution $q=0$, and we get the eigenfunctions
\begin{equation}\label{expliciteigf}
\phi_{j,1}=e^{(\lambda_{j}x+\lambda^3_{j}t)}, \quad \phi_{j,2}=e^{(-\lambda_{j}x-\lambda^3_{j}t)},  j=1,3, 5,\cdots , 2n-1.
\end{equation}

\begin{figure}
\mbox{\hspace{-0.6cm}}
\subfigure[]{\includegraphics[width=0.6\textwidth]{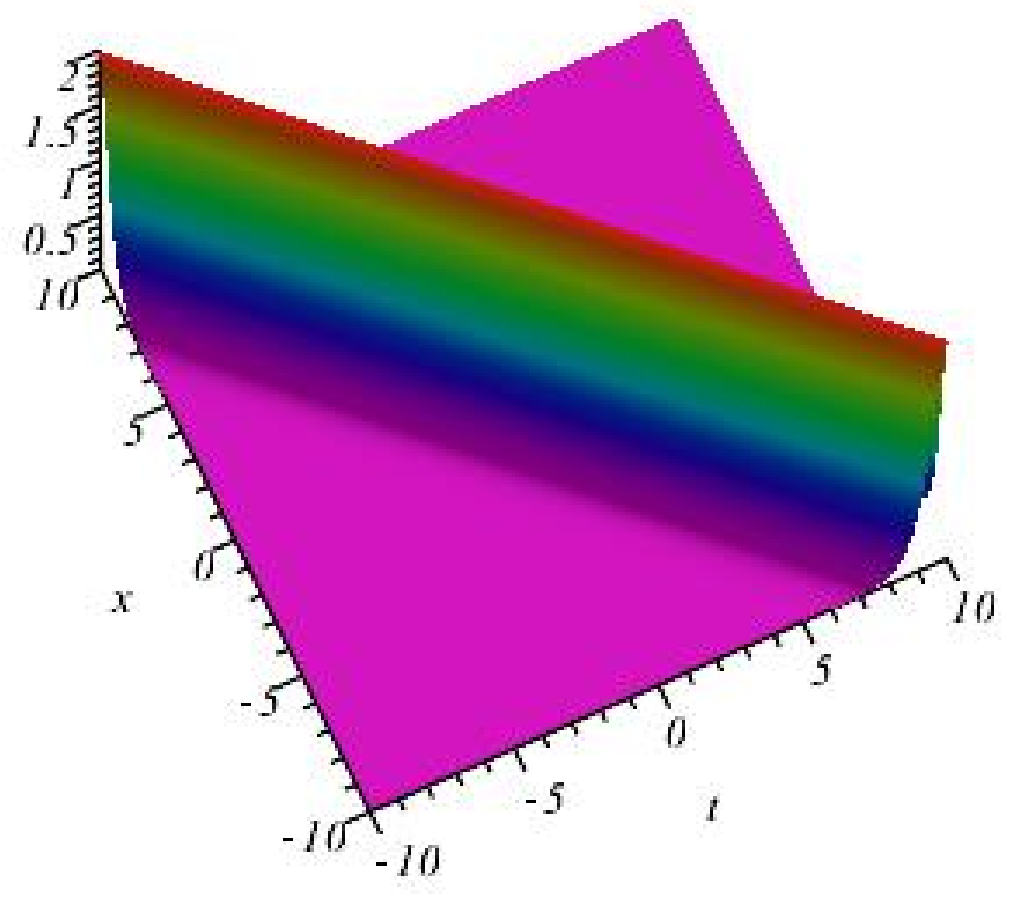}}\mbox{\hspace{-.5cm}}\subfigure[]{\includegraphics[width=0.45\textwidth]{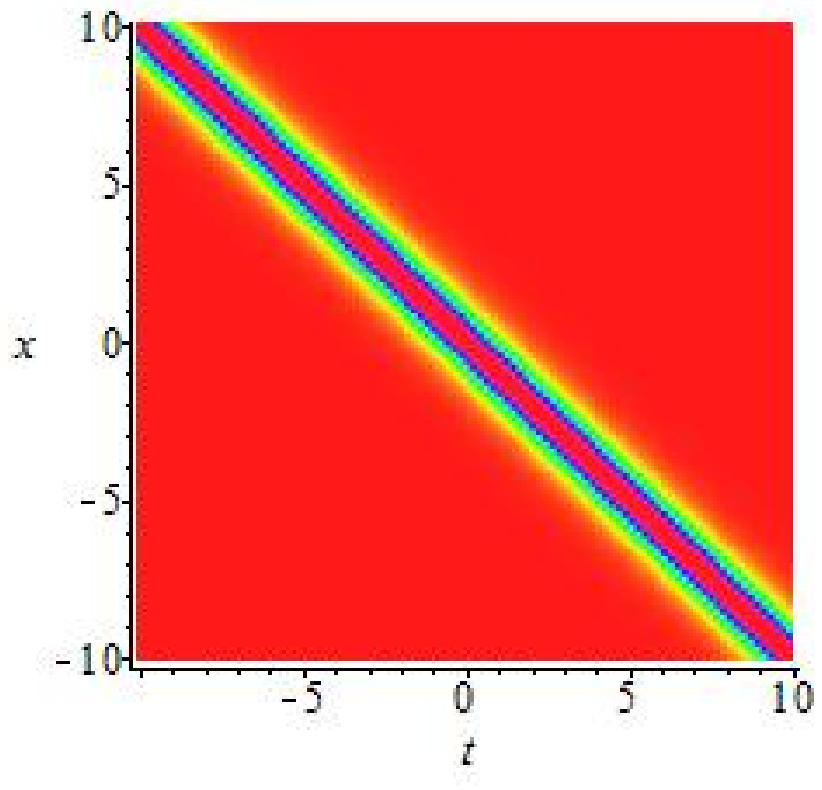}}
\caption{The evolution of a one-soliton solution $q^{[1]}(x,t)$ in the ($x,t$)-plane. Here the parameter is $\lambda_{1}=1$.
Panel (b) is a density plot of panel (a).}
\label{fig:1}       
\end{figure}

Taking the above eigenfunctions back into Eq. (\ref{13}) and using the reduction condition given by Eq. (\ref{reduction}),
then it yields an explicit form of the $n$-order soliton solution.
We set $n=1$ in equation (\ref{13}), then the single soliton solution is given by
\begin{equation}\label{ssoliton}
q^{[1]}=q_{1-s}(H)=\frac{4\lambda_1e^{2\lambda_1x+2\lambda_1^3t}}{1+e^{4\lambda_1x+4\lambda_1^3t}}=2\lambda_1 sech(2H), \quad
H=\lambda_1^3t+\lambda_1 x,
\end{equation}
which  is plotted in Fig. \ref{fig:1}.

If we set $n=2$,  the  Eq. (\ref{13})  generates a  two-soliton solution, namely
\begin{equation}
\mbox{\hspace{-0.05cm}}
q^{[2]}=\frac{4(-\lambda_{1}^2+\lambda_{3}^2)[\lambda_{1}(e^{-q}+e^{q})-\lambda_3(e^{-p}+e^{p})]}
{-(\lambda_1+\lambda_{3})^2(e^{p-q}+e^{-p+q})-(\lambda_1-\lambda_{3})^2(e^{-p-q}+e^{p+q})+8\lambda_1 \lambda_{3}}.
\end{equation}

\begin{figure*}
\mbox{\hspace{-0.6cm}}
\subfigure[]{\includegraphics[width=0.6\textwidth]{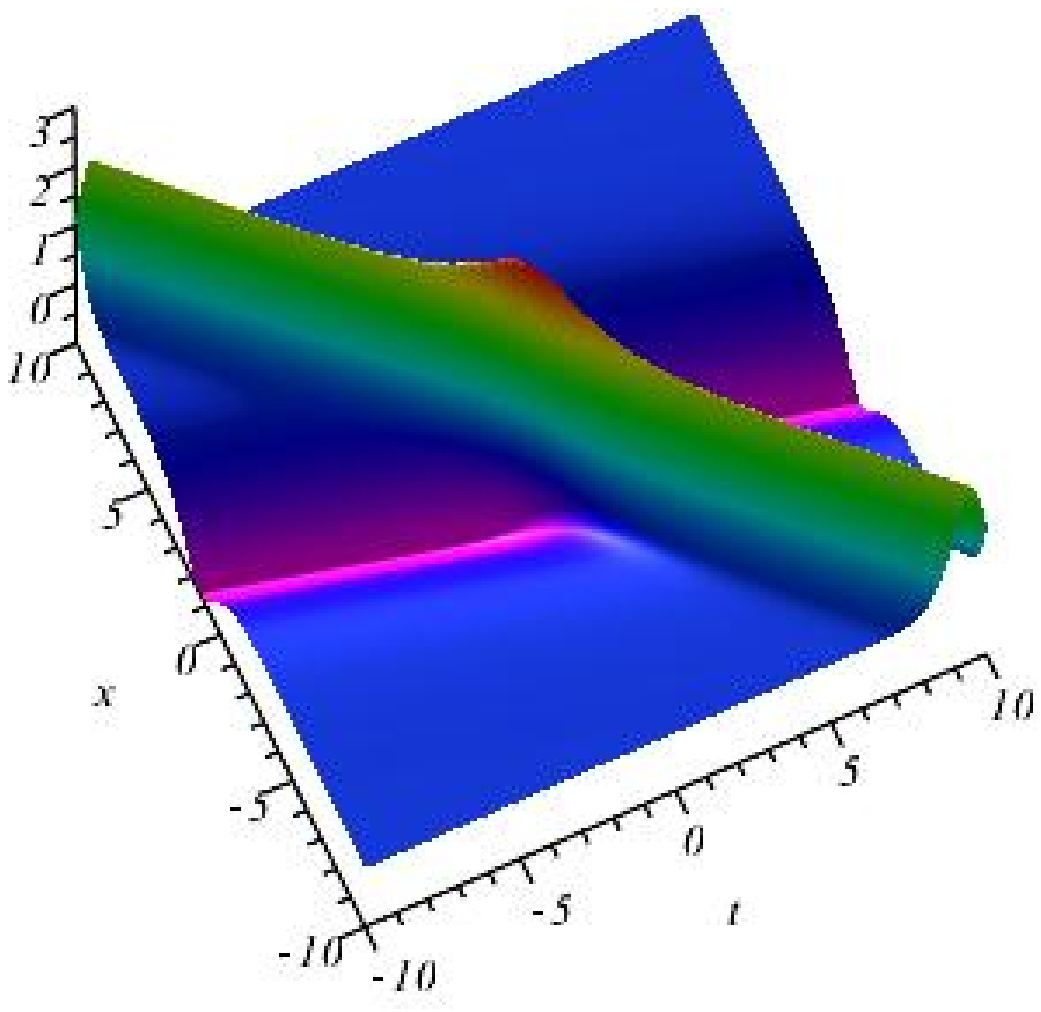}}\mbox{\hspace{-.5cm}}\subfigure[]{\includegraphics[width=0.45\textwidth]{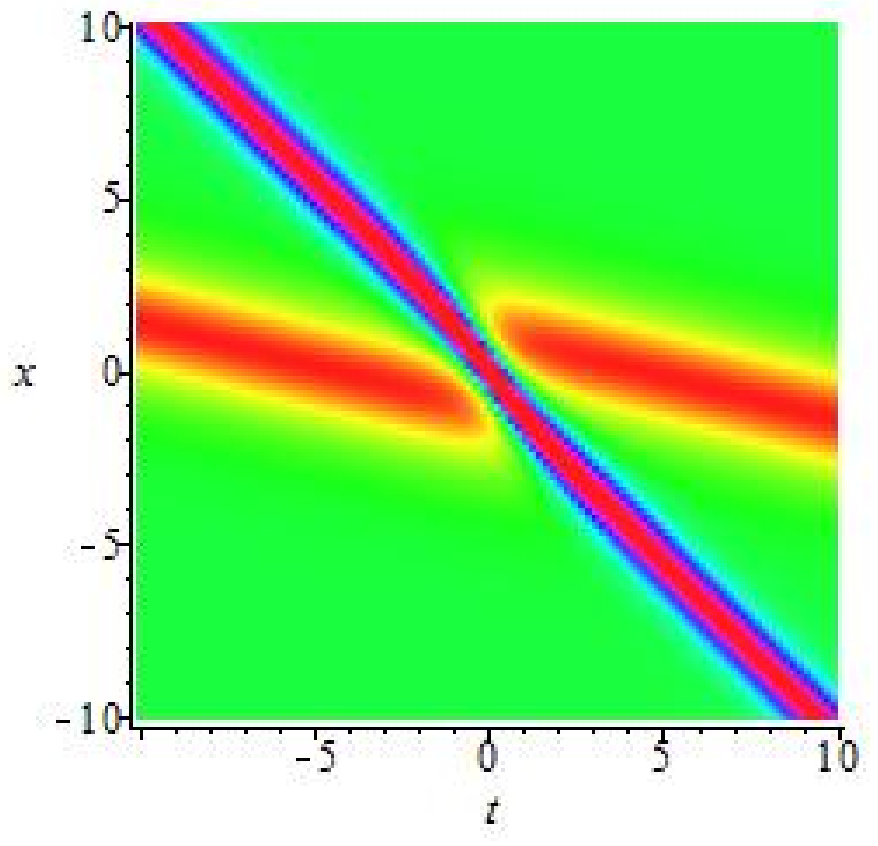}}
\caption{The evolution in the ($x,t$)-plane of a two-soliton solution $q^{[2]}(x,t)$ consisting of one bright soliton and one dark soliton. Here the parameters  are $\lambda_{1}=1$  and $\lambda_{3}=\frac{1}{2}$.
Panel (b) is a density plot of panel (a).}
\label{fig:2}       
\end{figure*}

\begin{figure*}
\mbox{\hspace{-0.6cm}}
\subfigure[]{\includegraphics[width=0.6\textwidth]{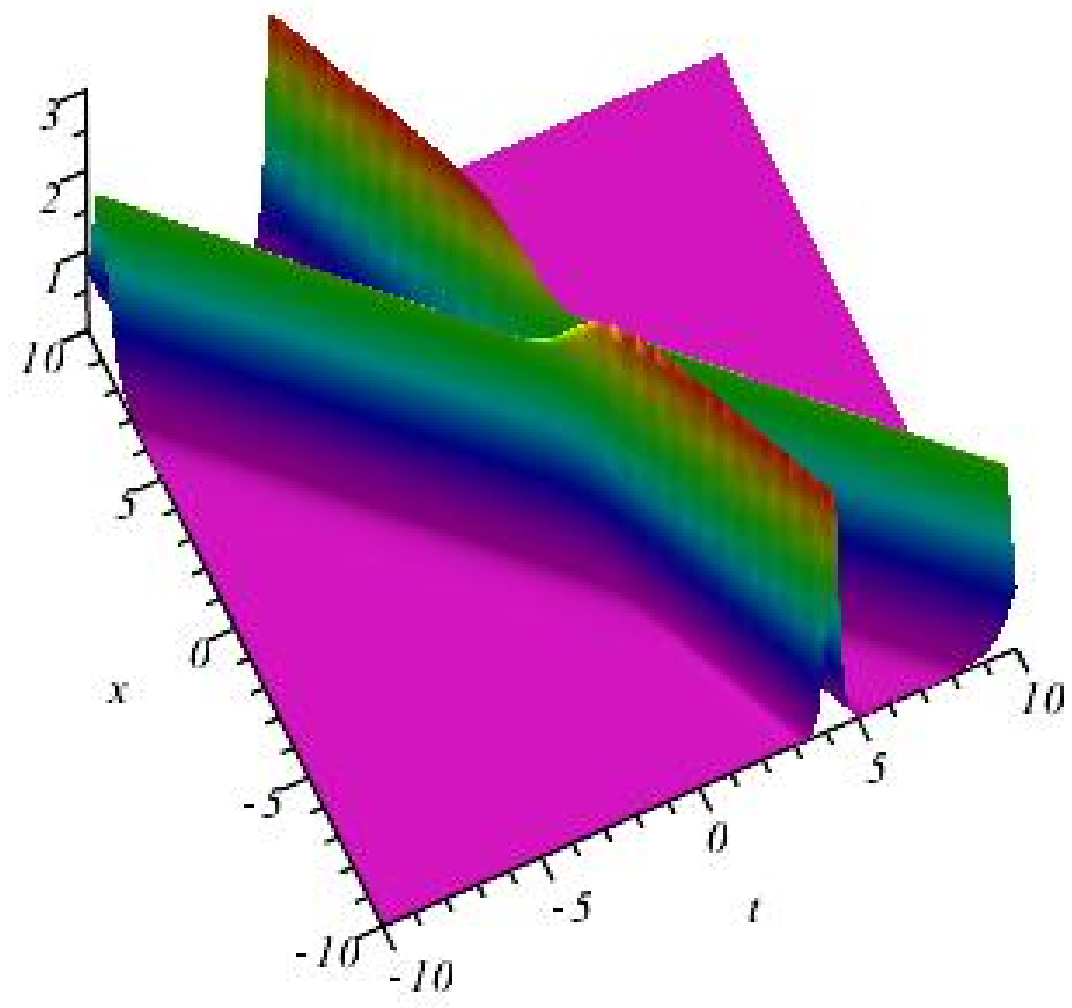}}\mbox{\hspace{-0.65cm}}\subfigure[]{\includegraphics[width=0.58\textwidth]{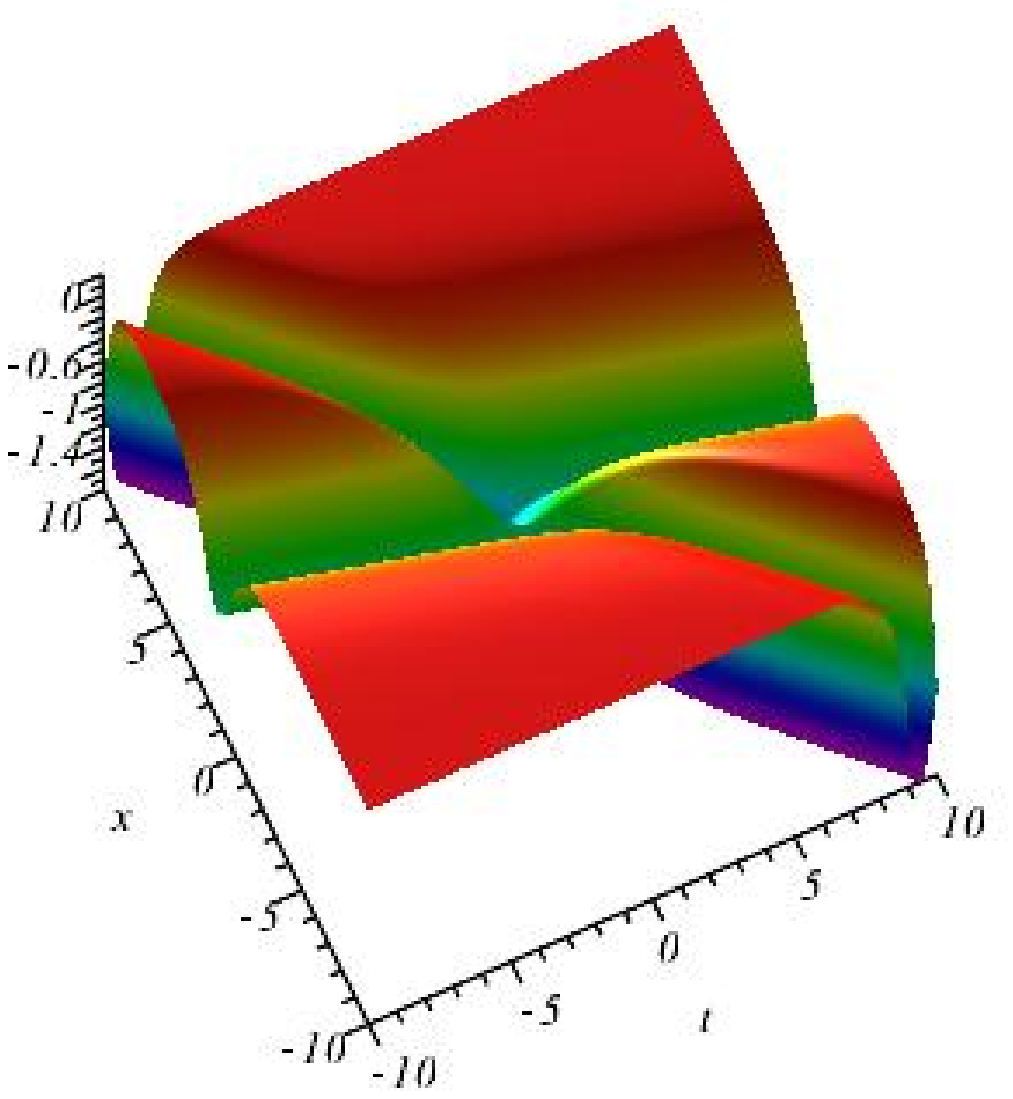}}
\caption{The evolution of other types of two-soliton solutions $q^{[2]}(x,t)$ in the ($x,t$)-plane. Panel (a) shows the evolution of two bright solitons  with parameters  $\lambda_{1}=\frac{3}{2}$ and $\lambda_{3}=-1$, and panel (b) diplays the evolution of two dark solitons  with parameters  $\lambda_{1}=-1$ and $\lambda_{3}=\frac{1}{2}$.
}
\label{fig:22}       
\end{figure*}

\begin{figure*}
\mbox{\hspace{-0.6cm}}
\subfigure[]{\includegraphics[width=0.6\textwidth]{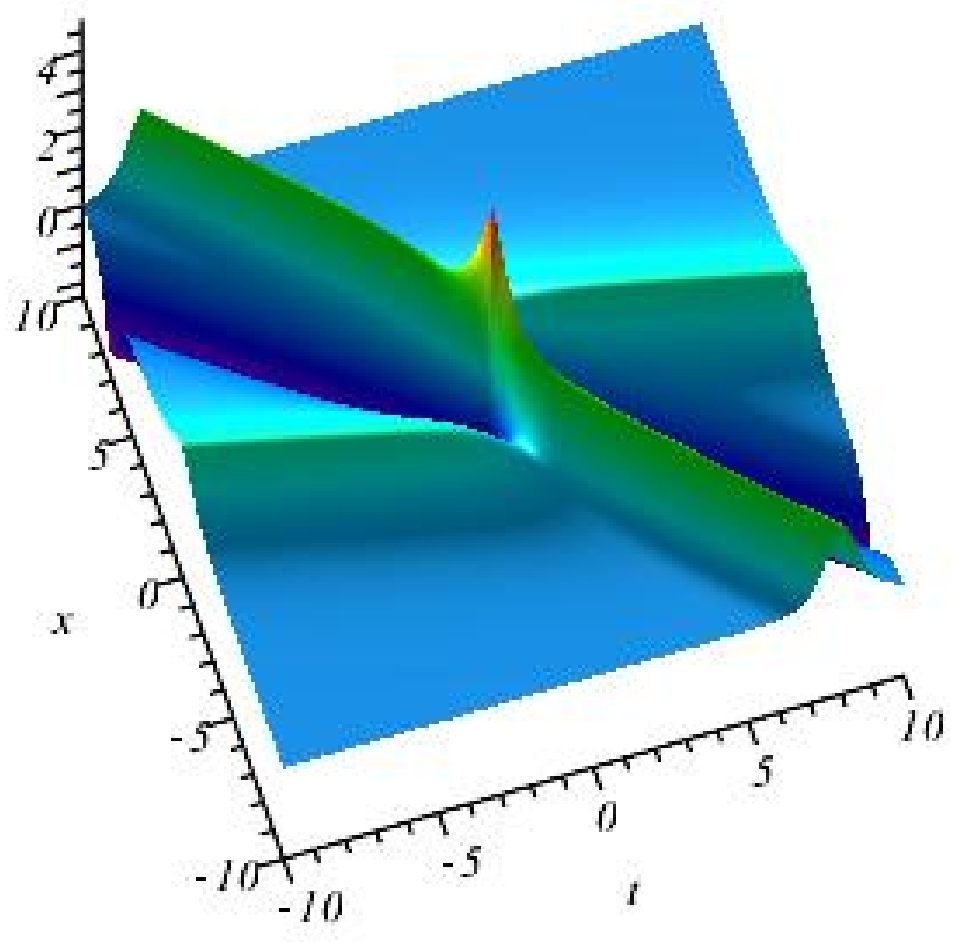}}\mbox{\hspace{-.6cm}}\subfigure[]{\includegraphics[width=0.45\textwidth]{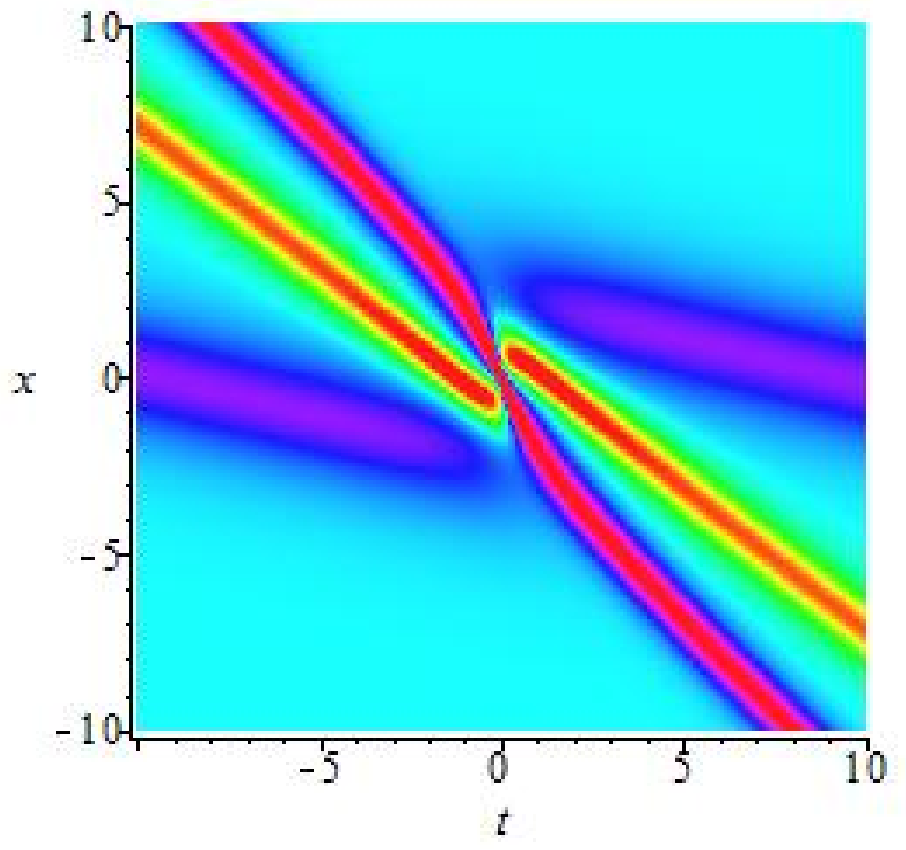}}
\caption{The evolution in the ($x,t$)-plane of a three-soliton solution $q^{[3]}(x,t)$  consisting of  two bright solitons and one dark soliton. The parameters  are $\lambda_{1}=1$, $\lambda_{3}=\frac{1}{2}$, and $\lambda_{5}=\frac{9}{10}$.
Panel (b) is a density plot of panel (a).}
\label{fig:3}       
\end{figure*}

\begin{figure*}
\mbox{\hspace{-.6cm}}
\subfigure[]{\includegraphics[width=0.6\textwidth]{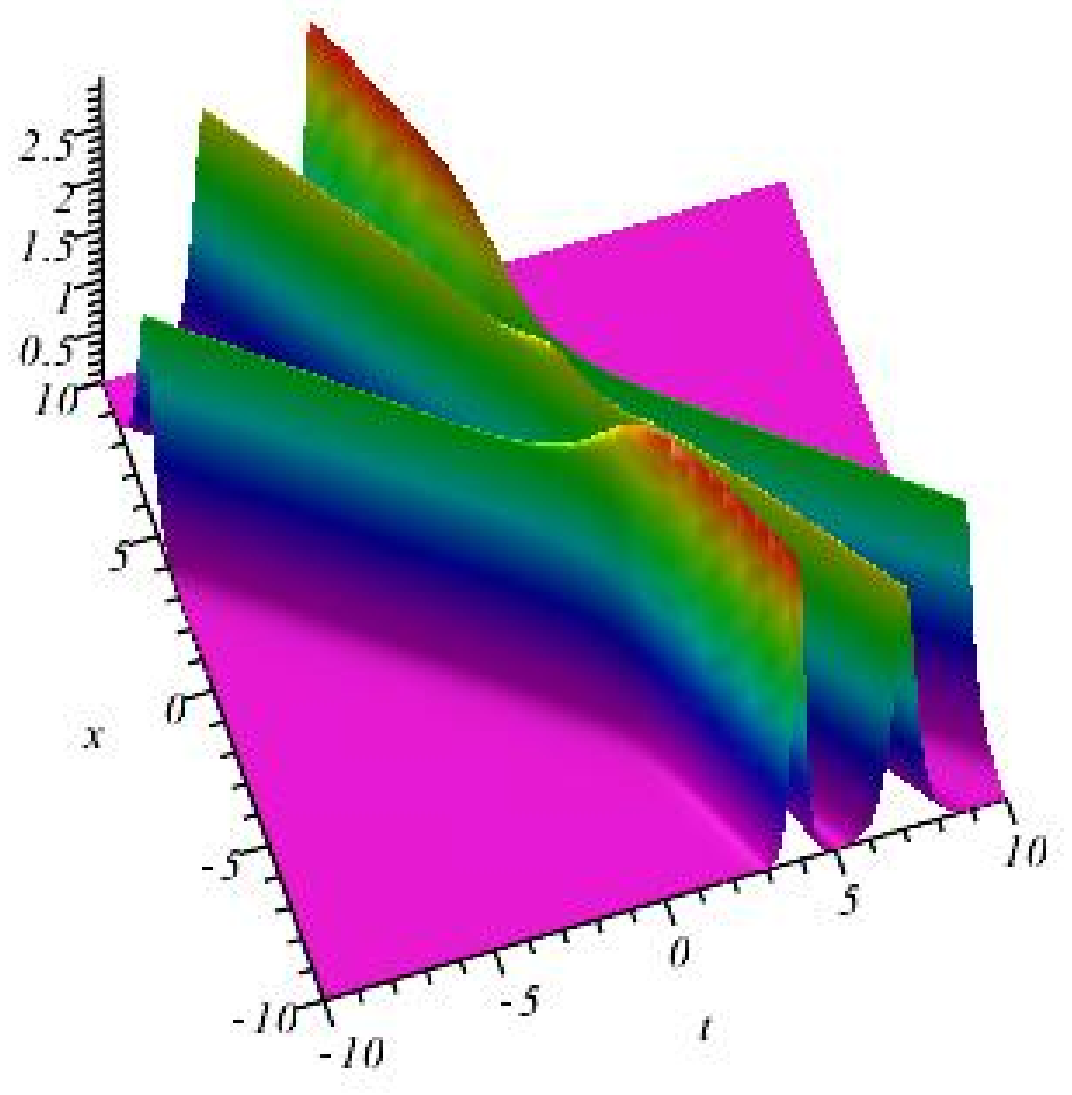}}\mbox{\hspace{-.8cm}}
\subfigure[]{\includegraphics[width=0.55\textwidth]{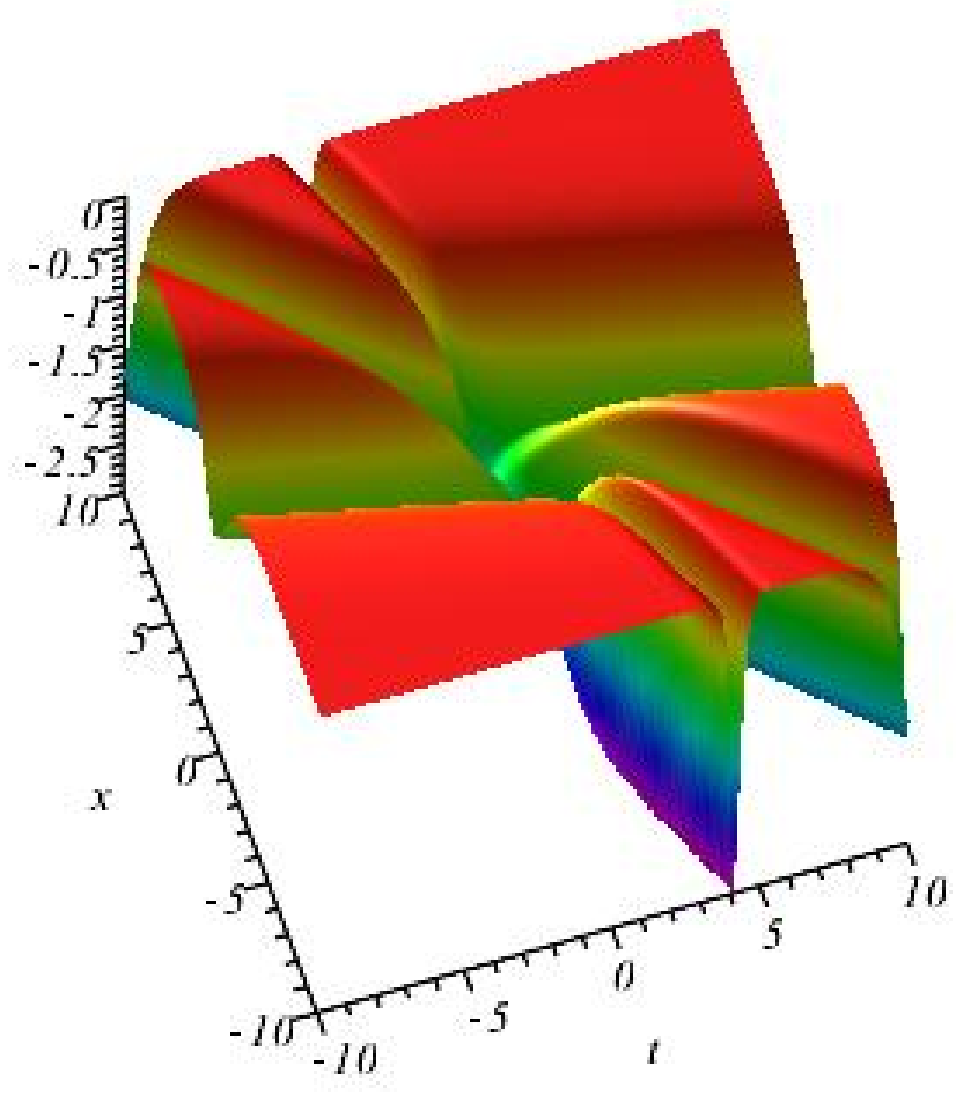}}
\caption{The evolution of other types of three-soliton solutions $q^{[3]}(x,t)$ in the ($x,t$)-plane.
 Panel (a) shows the evolution of three bright solitons  with parameters  $\lambda_{1}=\frac{9}{10}$, $\lambda_{3}=-\frac{6}{5}$, and $\lambda_{5}=\frac{3}{2}$,  and panel  (b) illustrates the evolution of three dark soliotns  with parameters  $\lambda_{1}=\frac{1}{2}$, $\lambda_{3}=1$, and $\lambda_{5}=-\frac{6}{5}$.
}
\label{fig:33}       
\end{figure*}
\noindent Here  $p=2H, \,  q=2(\lambda_{3}^3t+\lambda_{3}x)$. Three kinds of two-soliton solutions consisting of one bright soliton and one dark soliton, two bright solitons, and two dark solitons are  plotted in Fig. \ref{fig:2}, Fig. \ref{fig:22}(a) and Fig. \ref{fig:22}(b), respectively. Similarly,  three different cases of three-soliton solutions (see Eq. (\ref{13}) are  also plotted in Figs. \ref{fig:3} and \ref{fig:33}, namely two bright solitons and one dark soliton (see Fig. \ref{fig:3}),
 and  three bright solitons and three dark solitons (see Fig. \ref{fig:33}).

It is easy to see that the denominator in  the two-soliton solution is zero in the degenerate case when $\lambda_1=\lambda_3$. In general, the $n$-soliton
solution becomes an indeterminate form $\frac{0}{0}$ when $\lambda_j\rightarrow \lambda_1 \, ( j=1, 3, 5, \cdots, 2n-1)$.

However, if we set $\lambda_j=\lambda_1+\epsilon$ in the $n$-soliton solution, and then we perform the higher-order Taylor expansion (see, for example, Ref. \cite{Rj,Mu}), we then get the $n$-positon solution
\begin{equation}\label{npositon}
q_{n-p}=2\frac{N_{2n}'}{W_{2n}'},
\end{equation}
where
$$
N_{2n}'=(\frac{\partial^{n_{i}-1}}{\partial\epsilon^{n_{i}-1}}|_{\epsilon=0}(N_{2n})_{ij}(\lambda_{1}+\epsilon))_{2n\times2n},
$$
$$
W_{2n}'=(\frac{\partial^{n_{i}-1}}{\partial\epsilon^{n_{i}-1}}|_{\epsilon=0}(W_{2n})_{ij}(\lambda_{1}+\epsilon))_{2n\times2n},
$$
and $n_{i}=[\frac{i+1}{2}]$,${[i]}$ define the floor function of $i$.

\begin{figure}[!htpb]
\mbox{\hspace{-0.6cm}}
\centering
\subfigure[]{\includegraphics[width=0.6\textwidth]{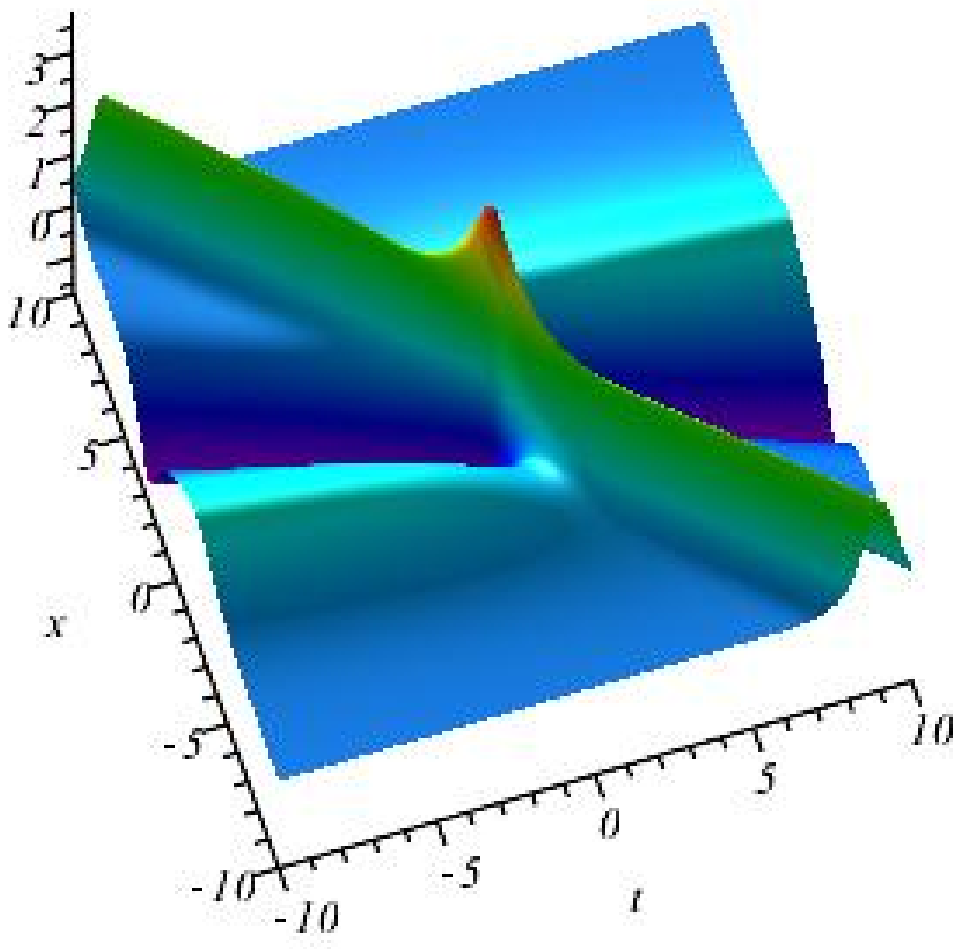}}\mbox{\hspace{-.5cm}}\subfigure[]{\includegraphics[width=0.45\textwidth]{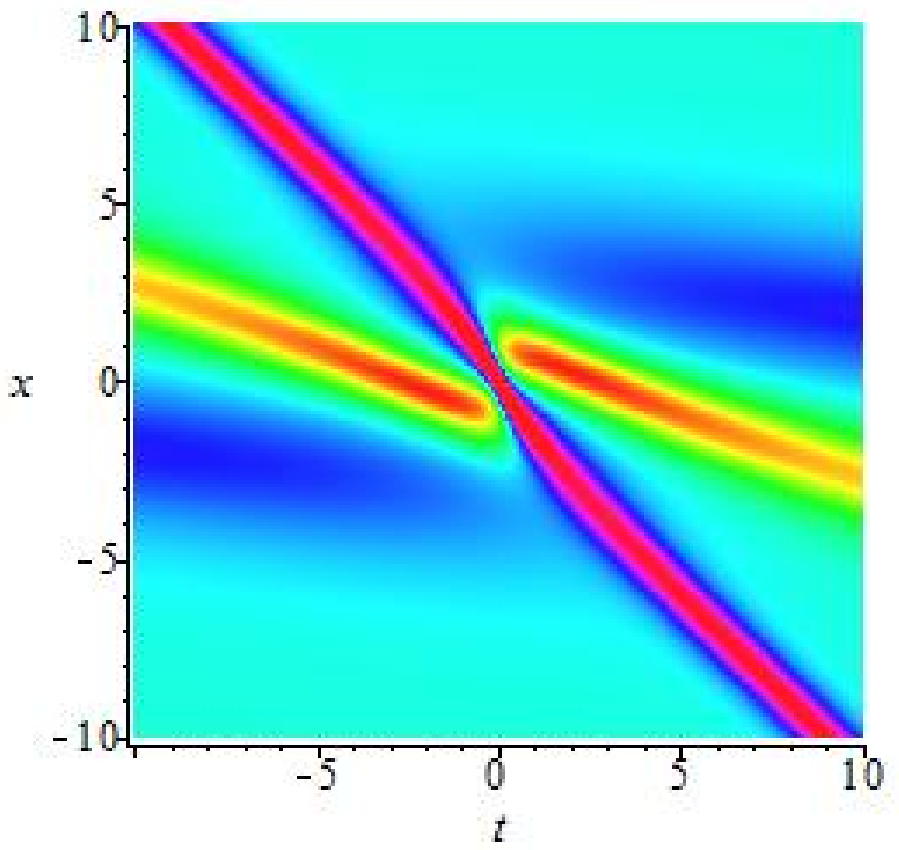}}
\subfigure[]{\includegraphics[width=0.5\textwidth]{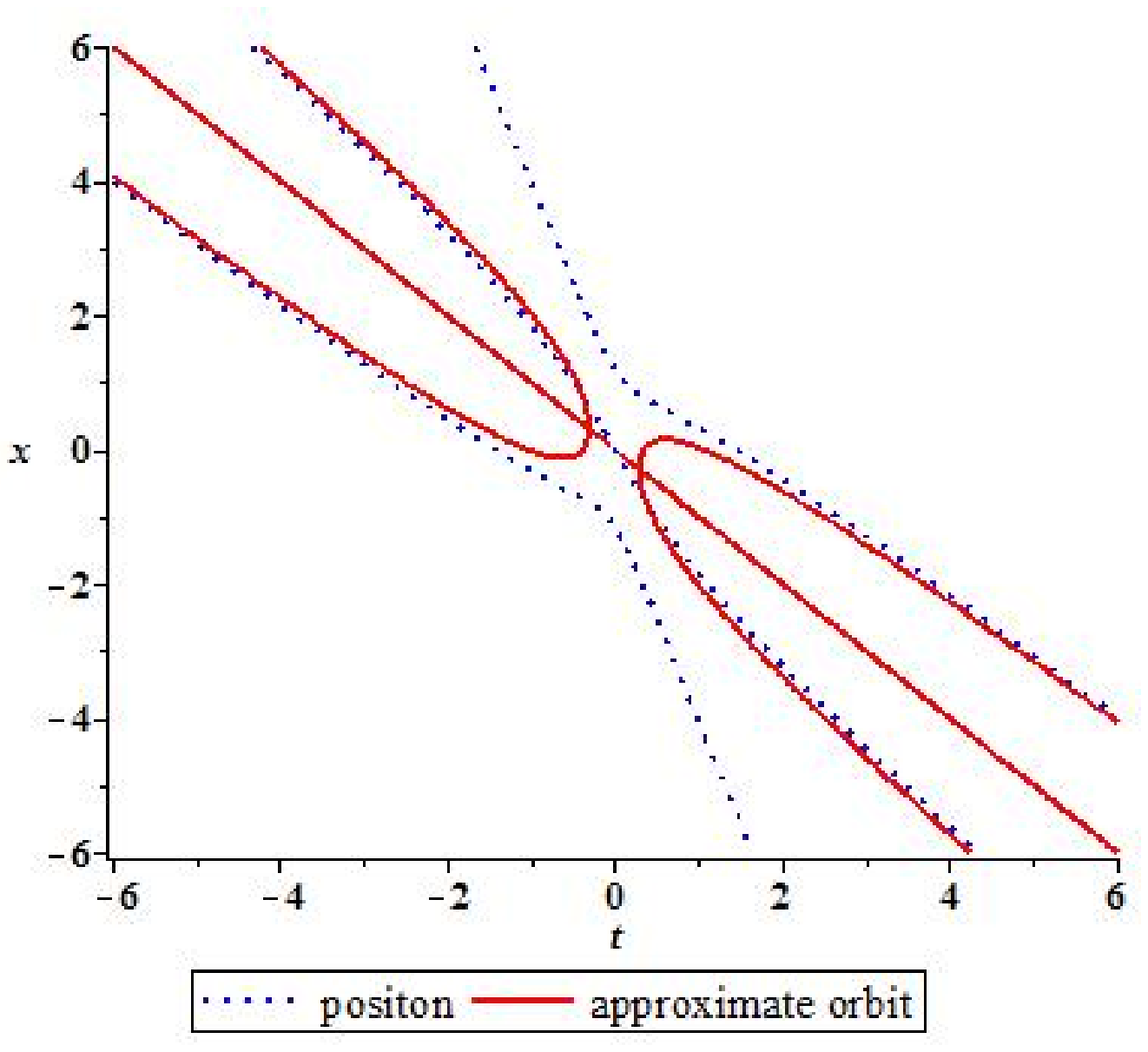}}
\caption{The evolution in the ($x,t$)-plane of a  two-positon solution $q_{2-p}(x,t)$ with the parameter $\lambda_{1}=1$.
Panel (b) is a  density plot of panel (a).  Panel (c) shows both the positon trajectories (the two outer blue dotted lines show the location of the positon maxima) and the approximate orbits (red solid lines) of the corresponding single-solitons obtained by the decomposition of the two-positon solution $q_{2-p}(x,t)$.  The middle red solid line corresponds to $H=x=t=0$.}
\label{fig:4}       
\end{figure}

\begin{figure*}
\mbox{\hspace{-0.6cm}}
\centering
\subfigure[]{\includegraphics[width=0.6\textwidth]{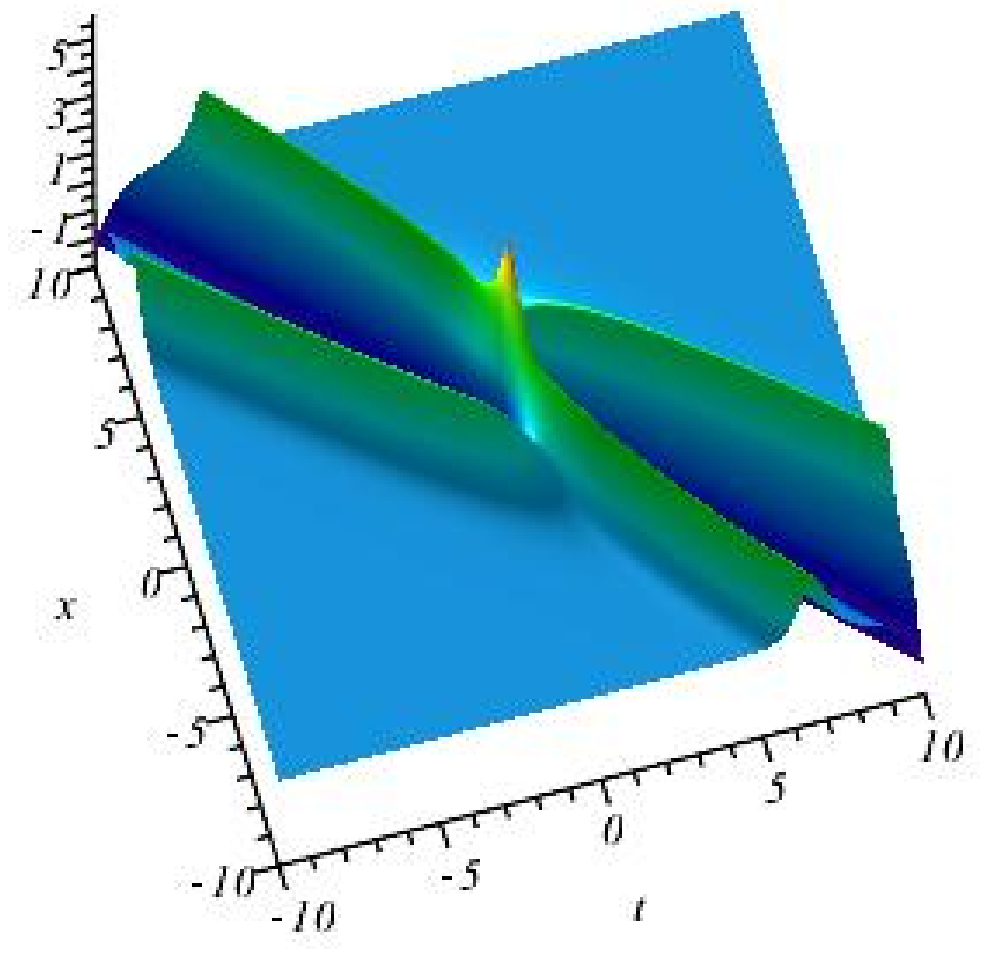}}\mbox{\hspace{-.5cm}}\subfigure[]{\includegraphics[width=0.45\textwidth]{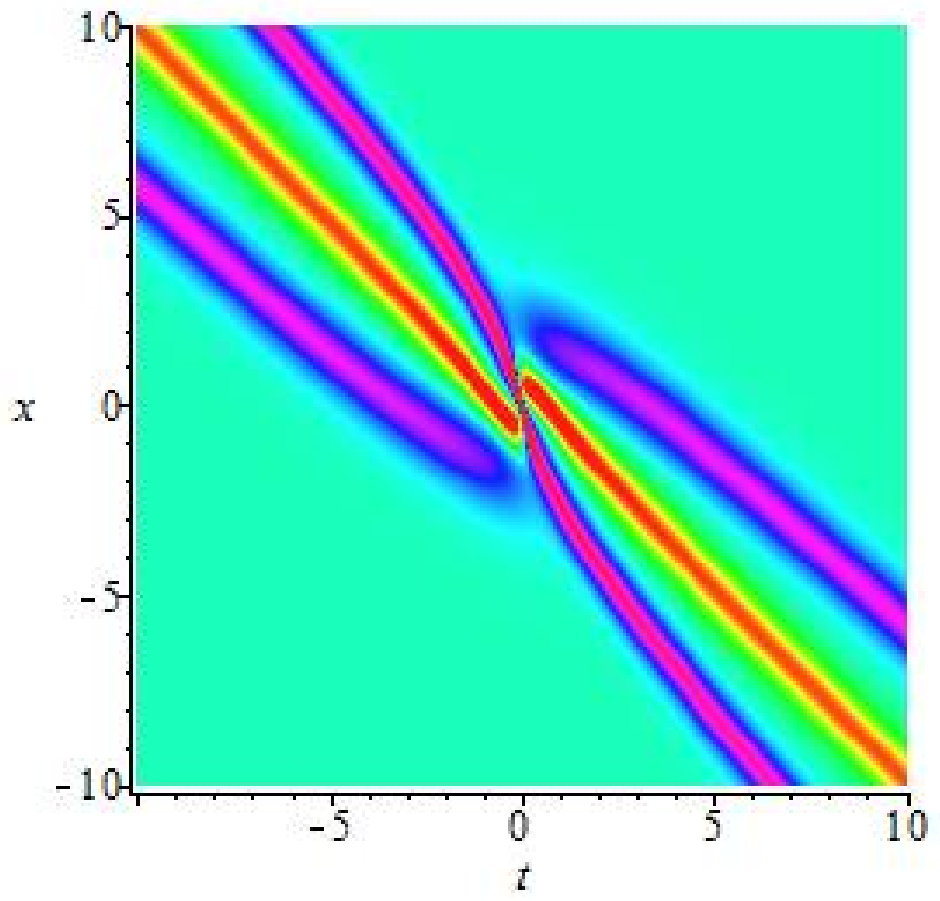}}
\subfigure[]{\includegraphics[width=0.5\textwidth]{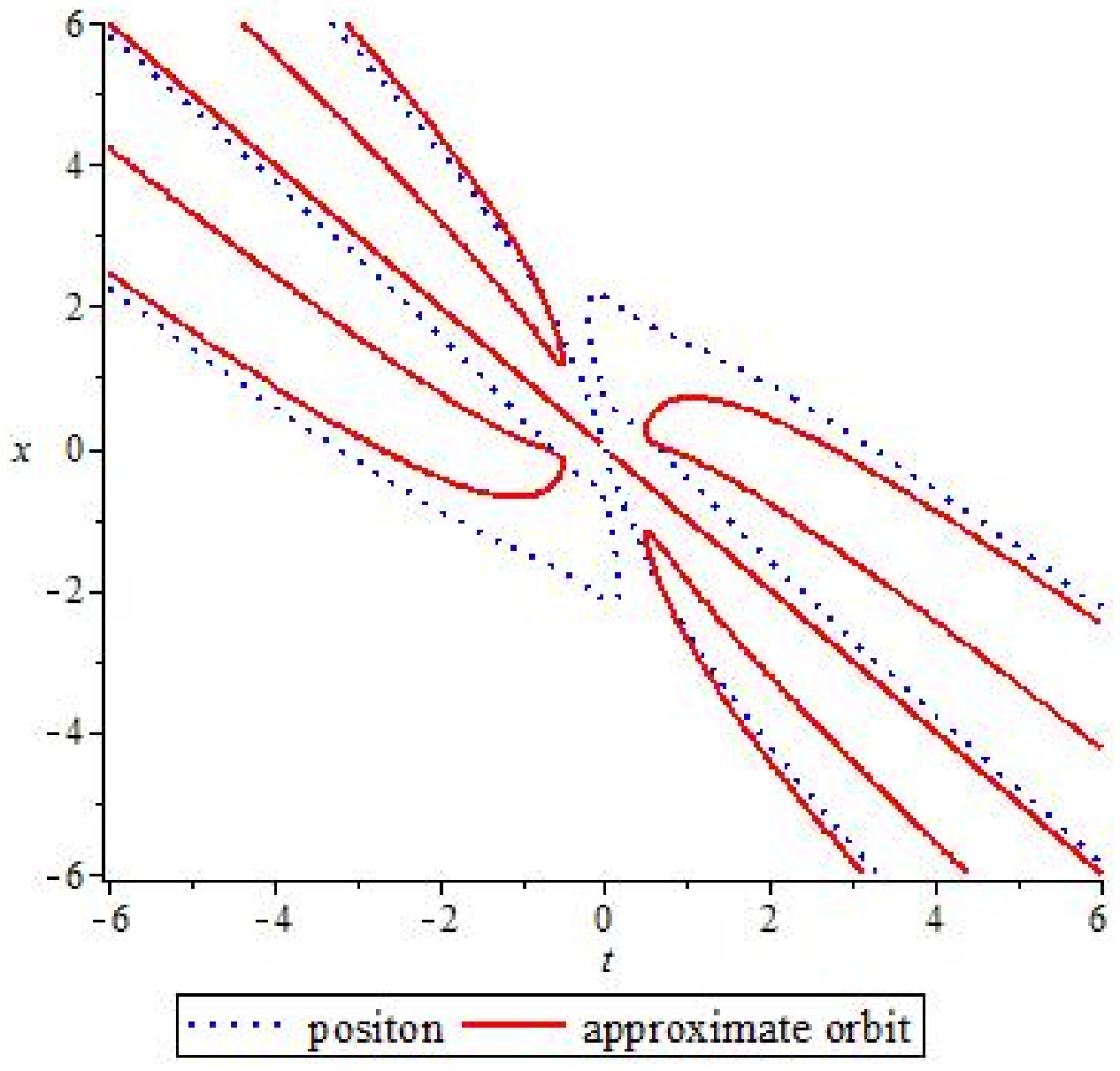}}
\caption{The evolution in the ($x,t$)-plane of a  three-positon  solution $q_{3-p}(x,t)$  with parameter $\lambda_{1}=1$.
Panel (b) is a  density plot of panel (a).  Panel (c) shows both the positon trajectories (the blue dotted lines) and the approximate orbits (red solid lines) of the corresponding single-solitons obtained by the decomposition of the three-positon solution $q_{3-p}(x,t)$.  The middle red solid line corresponds to $H=x=t=0$.}
\label{fig:5}       
\end{figure*}

If we set $n=2$,  the expression of two-positon solution is given by
\begin{equation}
q_{2-p}=\frac{8\lambda_{1}[(6\lambda_{1}^{3}t+2\lambda_{1}x-1)e^{2H}-(6\lambda_{1}^{3}t+2\lambda_{1}x+1)e^{-2H}]}{-\lambda_{1}^{2}(4x-12\lambda_{1}^{2}t)^2-4{\cosh^2(2H)}}.
\end{equation}
The typical two-positon solution is plotted in Fig. \ref{fig:4}. Similarly, we can get the three-positon solution from the three-soliton solution when $n=3$ in Eq. (\ref{npositon}) by performing the corresponding Taylor expansion as in the case of the two-positon solution. Because of the complexity of the exact form of the three-positon solution, we do not write down its explicit expression, but we have plotted it in \ref{fig:5}.

In what follows we will discuss the unique properties of the positon solutions. According to the expression of the two-positon solution, we see that the denominator of the $q_{2-p}$ cannot be zero, which proves the assertion that the positon solution of the focusing mKdV equation is nonsingular, a result that is completely different from the previous studies about the singular positon solution of the defocusing mKdV equation. At the same time, we can see from Fig. \ref{fig:4}(a) the smoothness of the positon solution. The above results inspires us to further study the key features of the positon solution of the focusing mKdV equation. We can easily see that the two-positon solution $q_{2-p}$ is not a travelling wave, and the trajectory of the positon $q_{2-p}$ is not a straight line, that is to say, it is a slowly changing curve. Then, we can explore the dynamics of smooth positons by looking at three main features: the decomposition procedure, the positon trajectories and the corresponding ``phase shifts''.


We will set $\lambda_{1}=1$ in order to simplify the calculations. The dynamics of smooth two-positon and three-positon solutions will be analyzed as follows:
\begin{itemize}
\item As is well known, a two-soliton solution can be decomposed into a sum of two single solitons with a phase shift at $|t|\gg 0$. The fact stimulates us consider a similar decomposition of the two-positon solution, because a smooth two-positon is obtained in a certain limit of the two-soliton solution solution. Naturally, we introduce the following decomposition of the two-positon solution
\begin{equation}\label{33}
q_{2-p}\approx q_{1-s}(H+c_{1})+q_{1-s}(H-c_{1}), \, H=(\lambda_1x+\lambda_1^3t)|_{\lambda_1=1}=x+t,
\end{equation}
when $|t|\longrightarrow\infty$

Here
\begin{equation}\label{34}
q_{1-s}(H+ c_{1})=\frac{4}{e^{2H+2c_{1}}+e^{-2H-2c_{1}}},
\end{equation}
\begin{equation}\label{35}
q_{1-s}(H- c_{1})=\frac{4}{e^{2H-2c_{1}}+e^{-2H+2c_{1}}},
\end{equation}
 where  $q_{1-s}(H)$ is the single-soliton solution given by  Eq. (\ref{ssoliton}). The phase shift $c_{1}$ can be determined by
 substituting  (\ref{34}) and (\ref{35}) into (\ref{33}) and   considering  the corresponding approximation  in the neighborhood of $H=0$.  It yields
\begin{equation}\label{36}
e^{4c_{1}}-2+e^{-4c_{1}}-64t^{2}\approx0.
\end{equation}
Note that the phase shift is a constant for the two-soliton solution.  However, the corresponding ``phase shift" $c_{1}$ for the positon solution is an undetermined function of $x$ and $t$.  By a simple calculation,  $c_{1}=-\frac{\ln(64t^2)}{4}$ is an approximate solution of Eq. (\ref{36}) for $|t|>> 1$.  Therefore a two-positon solution of the focusing  mKdV equation can be  decomposed as
\begin{equation}\label{32}
q_{2-p}\approx q_{1-s}(H+\frac{\ln(64t^2)}{4})+q_{1-s}(H-\frac{\ln(64t^2)}{4}),
\end{equation}
when $|t|\longrightarrow\infty$. Here, $H=x+t$, and very good approximate trajectories are two curves defined by $x+t\pm \frac{\ln(64t^2)}{4}=0$.  By comparing with the correponding decomposition of a multi-soliton solution into several single soliton solutions,  the appearance of the variable
``phase shift" in the decomposition of multi-positon solution into several single-soliton solutions is the key difference between the two decomposition processes.
\end{itemize}

\begin{itemize}
\item For a three-positon solution, we get
\begin{equation}\label{38}
q_{3-p}\approx q_{1-s}(H+c_{2})+q_{1-s}(H-c_{2})+q_{1-s}(H),
\end{equation}
when $|t|\longrightarrow\infty$ (here $H=(\lambda_1x+\lambda_1^3t)|_{\lambda_1=1}=x+t$ as before). Certainly, the phase shift $c_{2}$ still is an undetermined function of $x$ and $t$. Here we have $c_2=\frac{\ln 1024t^4}{4}$, and thus very good   approximate trajectories are the three curves defined by $x+t\pm \frac{\ln 1024t^4}{4}=0$ and $x+t=0$.
\end{itemize}

\section{Summary and discussion}
\label{sec:3}
In this paper, a determinant representation of the  $n$-fold Darboux transformation $T_{n}$ of the focusing real mKdV equation was given. Using this representation,  we have obtained the $n$-soliton solution $q^{[n]}$, and thus we explicitly provided  the  one-, two-, and three-soliton solutions.  Furthermore,  a general expression of the smooth, nonsingular  multi-positon solution of the focusing  mKdV equation was calculated by using a certain Taylor expansion in the corresponding determinant representation of the multi-soliton solution (see Eq. (\ref{npositon})).  Finally, we further analyzed the unique properties of the positon solution of the focusing  mKdV equation from three points of view: the decomposition procedure, the approximate trajectories, and the corresponding ``phase shifts". By comparing the case of the decomposition of the multi-soliton solution into single solitons with that of the decomposition of the multi-positon solution into single solitons we arrive at the interesting result that in the latter case the ``phase shifts" are variable and the approximate trajectories are not straight lines (see Figs. \ref{fig:4} and \ref{fig:5}). It is worth mentioning that Figs. \ref{fig:2} to \ref{fig:33} show that we cannot get pure bright or dark solitons when all the
eigenvalues of multi-soliton solutions are positive numbers or are negative ones. This facts implies naturally that
a multi-positon solution is a combination of several bright and dark solitons, see the two-positons shown in Fig.
\ref{fig:4} and the three-positons shown in Fig. \ref{fig:5}. This interesting result comes from the fact that the  multi-positon solution is obtained in the special limit when the eigenvalues $\lambda_j$ ($j=3, 5, \cdots$) tend to the same eigenvalue $\lambda_1$.

We may conclude that the obtained nonsingular multi-positon solutions of the focusing mKdV equation will undoubtedly be useful for further studies in this area and will give insights in modeling a diverse set of nonlinear wave phenomena in many relevant physical settings.

From our study it is clear that the  smooth $n$-positon solution  is  expressed by  a mixture of polynomial and hyperbolic functions,  similar to the multi-pole
solutions of the real and complex mKdV and the NLS equations, which were reported during the past three decades by using the classical inverse scattering and Hirota methods \cite{multi-pole mKdV}-\cite{focusing}. Comparing our results with the known results we would like to point out the following:
1) Equation (\ref{npositon}) provides a simple closed formula to calculate the $n$-positon solution;  2) Equations (\ref{32}) and (\ref{38}) provide
simple and direct expressions for the decomposition of multi-positon solutions into single solitons, and also show more precise formulas for the ``phase shifts"
and soliton trajectories. The corresponding soliton trajectories have been illustrated numerically in Figs. \ref{fig:4} and \ref{fig:5}. Moreover, it is worthy to further study the relationship between the smooth positon solutions and the multi-pole solutions for other physically relevant nonlinear evolution equations.

\begin{acknowledgements}
This work is supported by the NSF of China under Grant No. 11671219, and the K. C. Wong Magna Fund in  Ningbo University.  We thank members of our group at Ningbo University for
useful discussions on this manuscript.
\end{acknowledgements}



\end{document}